\documentclass[pre,twocolumn,aps,showpacs]{revtex4}
\usepackage{graphicx}

\begin{document}
\title{Surface magnetoinductive breathers in two-dimensional magnetic metamaterials}

\author{Maria Eleftheriou$^{1,2}$, Nikos Lazarides$^{3,4}$, George P.
Tsironis$^{3}$ and Yuri S. Kivshar$^{5}$
}
\affiliation{
$^{1}$Department of Materials Science and Technology, University of Crete,
P.O. Box 2208, Heraklion 71003, Crete,  Greece \\
$^{2}$Department of Music Technology and Acoustics, Technological
Educational Institute of Crete, Rethymno 74100, Crete, Greece \\
$^{3}$Department of Physics, University of Crete
and Institute of Electronic Structure and Laser
Foundation for Research and Technology-Hellas,
P.O. Box 2208, Heraklion 71003, Greece \\
$^{4}$Department of Electrical Engineering, Technological Educational
Institute of Crete, P.O. Box 140,  Heraklion 71500, Crete,
Greece\\
$^{5}$Nonlinear Physics Center, Research School of Physics and
Engineering, Australian National University, Canberra ACT 0200, Australia
}
\date{\today}

\begin{abstract}
We study discrete surface breathers in two-dimensional lattices of 
inductively-coupled split-ring resonators with capacitive nonlinearity. We 
consider both Hamiltonian and dissipative systems and analyze the properties 
of the modes localized in space and periodic in time (discrete breathers) 
located in the corners and at the edges of the lattice. We find that surface 
breathers in the Hamiltonian systems have lower energy than their bulk 
counterparts, and they are generally more stable.
\end{abstract}

\maketitle

Theoretical results on the existence of novel types of discrete
surface solitons localized in the corners or at the edges of
two-dimensional photonic lattices~\cite{makris_2D,pla_our,pre_2D}
have been recently confirmed by the experimental observation of
two-dimensional surface solitons in optically-induced photonic
lattices~\cite{prl_1} and two-dimensional waveguide arrays
laser-written in fused silica~\cite{prl_2,ol_szameit}. These
two-dimensional nonlinear surface modes demonstrate novel features
in comparison with their counterparts in truncated one-dimensional
waveguide arrays~\cite{OL_george,PRL_george,OL_molina}. In
particular, in a sharp contrast to one-dimensional
surface solitons, the mode threshold is lower at the surface than
in a bulk making the mode excitation easier~\cite{pla_our}.

Recently, it was shown~\cite{LTK} that, similar to discrete solitons
analyzed extensively  for optical systems,  surface discrete breathers
can be excited near the edge of  a one-dimensional metamaterial created
by a truncated array of nonlinear split-ring resonators.
Networks of split-ring resonators (SRRs) that have nonlinear capacitive elements
can support nonlinear localized modes or discrete breathers (DB's) under
rather general conditions that depend primarily on the inductive coupling
between SRRs and their resonant frequency~\cite{LET,ELT}. The corresponding
one-dimensional surface modes have somewhat lower energy (in the
Hamiltonian case) and can easily be generated in one-dimensional SRR lattices~\cite{LTK}.

In this Brief  Communication, we develop further those ideas and analyze two-dimensional
lattices of split-ring resonators. Similar to the optical systems, we find
that two-dimensional lattices of inductively-coupled split-ring resonators
with capacitive nonlinearity can support the existence of long-lived two-dimensional
discrete breathers localized in the corners or at the edge of the lattice.

\begin{figure}[!ht]
\includegraphics[angle=0, width=0.8 \linewidth]{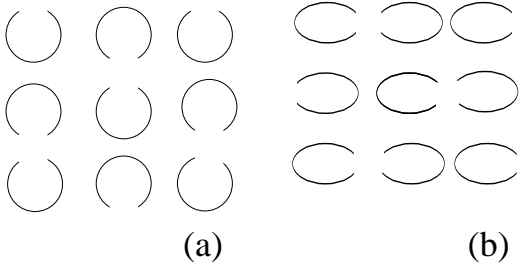}
\caption{Schematic of a two-dimensional lattice of split-ring resonators
for (a) planar and (b) planar-axial geometries.  In both the geometries
the magnetic component of the applied field is directed along the SRR axes,
while the electric field component is transverse to the slits.  }
\label{Fig1}
\end{figure}

We consider a two-dimensional lattice of SRRs in both planar and
planar-axial configuration [see Figs.~\ref{Fig1}(a,b)]. In the planar configuration,
all SRR loops are in the same plane with their centers forming an orthogonal lattice,
while in the planar-axial configuration the loops have a planar arrangement in one
direction and an axial configuration in the other direction. Each SRR  is
equivalent to a nonlinear RLC circuit, with an ohmic resistance $R$,
self-inductance $L$, and  capacitance $C$. We assume that the capacitor $C$
contains a nonlinear Kerr-type dielectric, so that the permittivity $\epsilon$
can be presented in the form,
\begin{eqnarray}
\label{0}
   \epsilon (|{\bf E}|^2) = \epsilon_0 \left( \epsilon_\ell + \alpha
      \frac{|{\bf E}|^2}{E_c^2} \right) ,
\end{eqnarray}
where ${\bf E}$ is the electric field with the characteristic value $E_c$,
$\epsilon_\ell$ is linear permittivity, $\epsilon_0$ is the
permittivity of the vacuum, and $ \alpha=+1~~(-1)$ corresponding to
self-focusing (self-defocusing) nonlinearity, respectively. As a result,
each SRR acquires the field-dependent capacitance $C(|{\bf E}|^2) = \epsilon ( |{\bf E}_g|^2 )\,
A/d_g$, where $A$ is the area of the cross-section of the SRR wire, ${\bf
E}_g$ is the electric field induced along the SRR slit, and $d_g$ is the size
of the slit. The field ${\bf E}_g$ is induced by the magnetic and/or
electric component of the applied electromagnetic field, depending on the relative
orientation of the field with respect to the SRR plane and the slits~\cite{Shardivov}.
Below we assume that  the magnetic component of the incident (applied) electromagnetic field is always perpendicular to the SRR plane, so that the electric field component is transverse to the slit.  With this assumption,
only the magnetic component of the field excites an electromotive force in SRRs, resulting
in an oscillating current in each SRR loop.  This results in the development of an oscillating voltage
difference $U$ across the slits or, equivalently, of an oscillating electric
field ${\bf E}_g$ in the slits.

If $Q$ is a charge stored in teach SRR capacitor, from a general
relation of a voltage-dependent capacitance $C(U)=dQ / dU$ and
Eq.~(\ref{0}), we obtain
\begin{eqnarray}
\label{1}
   Q = C_\ell
     \left( 1 + \alpha \frac{U^2}{3 \epsilon_\ell \, U_{c}^2}
     \right) U,
\end{eqnarray}
where $U =d_g E_g$, $C_\ell = \epsilon_0 \epsilon_\ell (A / d_g)$ is the linear capacitance,
and $U_{c} =d_g E_{c}$.
We assume that the arrays are placed in a time-varying and spatially uniform
magnetic field of the form
\begin{eqnarray}
\label{01}
  H = H_0 \, \cos(\omega t) ,
\end{eqnarray}
where $H_0$ is the field amplitude, $\omega$ is the field frequency,
and $t$ is the time variable. The excited electromotive force ${\cal E}$ ,
which is the same in all SRRs, is given by the expression
\begin{eqnarray}
\label{02}
  {\cal E} = {\cal E}_0 \, \sin(\omega t),
  \qquad {\cal E}_0 \equiv \mu_0 \, \omega \, S \, H_0 ,
\end{eqnarray}
where $S$ is the area of each SRR loop, and $\mu_0$ the permittivity of the
vacuum.  Each SRR exposed to the external field given by Eq. (\ref{01}) is a
nonlinear oscillator exhibiting a resonant magnetic response at a particular frequency
which is very close to its linear
resonance frequency $\omega_\ell = 1/ \sqrt{L \, C_{\ell}}$ (for $R\simeq 0$).

All SRRs in an array are coupled together due to magnetic dipole-dipole
interaction through their mutual inductances. However, we assume below only
the nearest-neighbor interactions, so that neighboring SRRs are coupled through
their mutual inductances $M_x$ and $M_y$. This is a good approximation in
the planar configurations [see Fig.\ref{Fig1}(a)], even if SRRs are located very
close. Validity of the nearest-neighbor approximation for the planar-axial configuration
[see Fig.\ref{Fig1}(b)] has been verified by taking into account the interaction of
SRRs with their four nearest neighbors. Assumimg that the mutual inductance 
$M_{x,y}^{(s)}$ between an SRR and its $s-$th neighbor decays with distance as $M_{x,y}^{(s)} \simeq M_{x,y} / s^3$~ \cite{ELT}, we find practically the same results.
Therefore, the electrical equivalent of
an SRR array in an alternating magnetic field is an array of nonlinear RLC
oscillators coupled with their nearest neighbors through their mutual
inductances; the latter  are being driven by identical alternating voltage sources.
Equations describing the dynamics of the charge $Q_{n,m}$ and
the current $I_{n,m}$ circulating in the $n,m-$th SRR may be derived
from Kirchhoff's voltage law for each SRR~\cite{LET,Shardivov}
\begin{eqnarray}
\label{2}
  \frac{dQ_{n,m}}{dt} &=& I_{n,m}  \\
\label{3}
    L \frac{dI_{n,m}}{dt} &+& R I_{n,m} + f (Q_{n,m})=
    \nonumber \\
   &-& M_{x} \left(\frac{dI_{n-1,m}}{dt}+\frac{dI_{n+1,m}}{dt} \right)
    \nonumber \\
   &-& M_{y} \left(\frac{dI_{n,m-1}}{dt}+\frac{dI_{n,m+1}}{dt} \right)
 	+ {\cal E},
\end{eqnarray}
where  $f(Q_{n,m})=U_{n,m}$ is given implicitly from Eq. (\ref{1}).
Using the relations
\begin{eqnarray}
\label{4}
  \omega_\ell^{-2} &=& L  C_\ell,~~\tau=t  \omega_\ell,
  ~~I_c = U_c  \omega_\ell  C_\ell,~~Q_c=C_\ell  U_c \\
  \label{5}
 {\cal E} &=& U_c \varepsilon,~~I_{n,m}=I_c i_{n,m},~~Q_{n,m} = Q_c q_{n,m} ,
\end{eqnarray}
and Eq. (\ref{02}), we normalize Eqs. (\ref{2}) and  (\ref{3}) to the form
\begin{eqnarray}
\label{6}
   \frac{d q_{n,m}}{d\tau} &=& {i_{n,m}} \\
\label{7}
  \frac{d i_{n,m}}{d\tau} &+&\gamma \, i_{n,m} + f (q_{n,m}) +
   \lambda_{x} \left( \frac{d i_{n-1,m}}{d\tau} +\frac{d i_{n+1,m}}{d\tau}
        \right)
	\nonumber \\
  &+&\lambda_{y} \left( \frac{d i_{n,m-1}}{d\tau} +\frac{d i_{n,m+1}}{d\tau}
        \right)
    = \varepsilon_0 \, \sin(\Omega\tau) ,
\end{eqnarray}
where $\gamma=RC_{\ell}\omega_{\ell}$ is the loss coefficient, $\lambda_{x,y}
=M_{x,y} / L$ are the the coupling parameters in the $x-$ and $y-$direction,
respectively, and $\varepsilon_0 = {\cal E}_0 / U_c$.  Note that the loss
coefficient $\gamma$, which is usually small ($\gamma \ll 1$), may
account both for Ohmic and radiative losses \cite{Kourakis}.  Neglecting
losses and without applied field, Eqs. (\ref{6}) and (\ref{7}) can be derived
from the Hamiltonian
\begin{eqnarray}
 \label{8}
  {\cal H} &=& \sum_{n,m} \left\{
      \frac{1}{2} \dot{q}_{n,m}^2  + V_{n,m} \right\}
      \nonumber \\
      &-&\sum_{n,m} \left\{
     \lambda_x \, \dot{q}_{n,m}\, \dot{q}_{n+1,m}
    +\lambda_y \, \dot{q}_{n,m}\, \dot{q}_{n,m+1}
      \right\}  ,
\end{eqnarray}
where the nonlinear on-site potential $V_{n,m}$ is given by
\begin{eqnarray}
 \label{9}
   V_{n,m} \equiv V( q_{n,m} ) =\int_0^{q_{n,m}} f(q_{n,m}') \, dq_{n,m}' ,
\end{eqnarray}
and $\dot{q}_{n,m} \equiv d{q}_{n,m} / d\tau$.
Analytical solution of Eq. (\ref{1}) for $u_{n,m}=f(q_{n,m})$
with the conditions of $u_{n,m}$ being real and
$u_{n,m}(q_{n,m}=0)=0$, gives the approximate expression
\begin{eqnarray}
\label{10}
   f(q_{n,m}) \simeq q_{n,m} -\frac{\alpha}{3\epsilon_\ell}q_{n,m}^{3}
     + 3\left(\frac{\alpha}{3\epsilon_\ell}\right)^{2} q_{n,m}^{5} ,
\end{eqnarray}
which is valid for relatively low $q_n$ ($q_n < 1,~~n=1,2,...,N$). Thus, the
on-site potential is soft for $\alpha=+1$ and hard for $\alpha=-1$.  In the 2D
case the mutual inductances $M_{x}$ and $M_{y}$ may differ both in their sign,
depending on the configuration, and their magnitude.  Actually, even in the
planar 2D configuration with $d_x=d_y$ a small anisotropy should be expected
because we consider SRRs having only one slit.  This anisotropy can be
effectively taken into account by considering slightly different coupling
parameters $\lambda_x$ and $\lambda_y$.  The coupling parameters
$\lambda_{x,y}$ as well as the loss coefficient $\gamma$ can be calculated
numerically for this specific model with high accuracy. However, for our
purposes, it is sufficient to estimate these parameters for realistic
(experimental) array parameters, ignoring the nonlinearity of the SRRs and the
effects due to the weak coupling as in Refs. \cite{LET,ELT}
with the following typical values  $\lambda \approx 0.02$ and $\gamma\approx 0.01$.

\begin{figure}[!ht]
\includegraphics[angle=0, width=1 \linewidth]{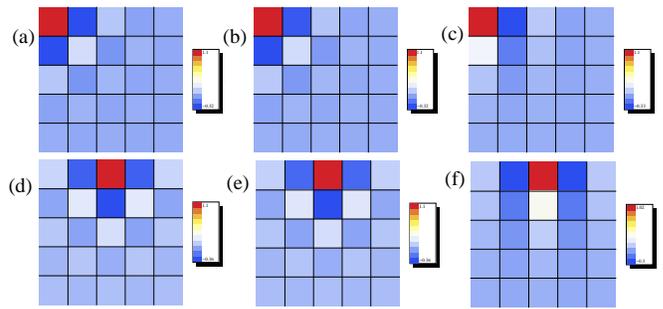}
\caption{Density amplitudes $q_{n,m}$  for discrete Hamiltonian
breathers constructed in (a-c) upper left corner  or (d-f)
upper edge of the lattice of $15\times 15$ sites, $\alpha=+1$ and 
$T_{b}=6.69$. (a,c) $\lambda_x=\lambda_y=0.028$, (b,e) $\lambda_{x}=0.026$ 
and $\lambda_{y}=0.029$, (c,f) $\lambda_{x}=0.029$ and $\lambda_{y}=-0.026$. 
All plots depict a $5\times 5$ sublattice that includes the breather zones.
}
\label{Fig2}
\end{figure}

\begin{figure}[!ht]
\includegraphics[angle=0, width=1 \linewidth]{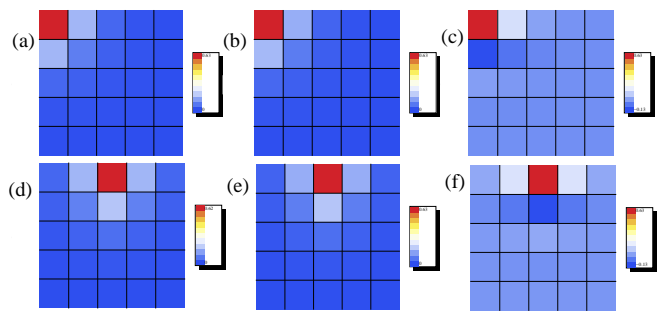}
\caption{Density amplitude $q_{n,m}$ for discrete Hamiltonian
breathers constructed in (a-c) upper left corner
or in (d-f) upper edge of a lattice of
$15\times 15$ sites, $\alpha=-1$ and $T_{b}=5.8$. (a,d)
$\lambda_x=\lambda_y=0.030$, (b,e) $\lambda_{x}=0.028$ and $\lambda_{y}=0.031$,
(c,f)  $\lambda_{x}=0.028$ and $\lambda_{y}=-0.025$. All plots depict the
$5\times 5$ sublattice around the linearly stable breathers.
}
\label{Fig3}
\end{figure}

We construct discrete breathers located in the corner of a two-dimensional lattice of $15\times 15$
sites using the anti-continuous limit method as in Ref. ~\cite{LET} for the set of Eqs. (\ref{6})-(\ref{7}),
setting  $\gamma=0$ and $\varepsilon_0=0$ (Hamiltonian discrete breathers).
For the case of $\alpha=+1$ corresponding to self-focusing nonlinearity and period
$T_{b}=6.69$, we may construct linearly stable breathers for parameters up to $\lambda_x=\lambda_y=0.029$.
Breather stability has been  checked through the Floquet monodromy matrix throughout the
paper. For the case where an anisotropy is introduced, $\lambda_x<\lambda_y$,
linearly stable discrete breathers can be constructed up to $\lambda_{x}=0.028$ and
simultaneously $\lambda_{y}=0.031$, or for the case of planar-axial
configuration up to $\lambda_{x}=0.031$ and  at the same time
$\lambda_{y}=-0.028$. If we look for discrete breathers
constructed in the middle of the upper edge of the lattice for example,
we find that the values of the coupling where an instability
occurs are slightly decreased (e.g. the upper stability limit of coupling for
planar geometry is $\lambda_x=\lambda_y=0.028$). Several cases of
linearly stable discrete breathers are shown in Fig.~\ref{Fig2} for $\alpha=+1$. The
same analysis holds for $\alpha=-1$ (defocusing nonlinearity) where the
upper stability limit for the values of
couplings are of the same order of magnitude as for $\alpha=+1$, both
for the corner and edge breathers (see Fig.~\ref{Fig3}). The breather period in the
latter case is $T_{b}=5.8$.

\begin{figure}[!ht]
\includegraphics[angle=0, width=1. \linewidth]{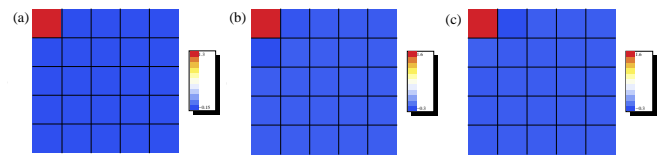}
\caption{Density amplitude $q_{n,m}$  for discrete dissipative
breathers for $\gamma=0.01$, $\varepsilon_0=0.04$, $\alpha=+1$ 
and $T_{b}=6.82$, constructed in the upper left corner for 
(a) $\lambda_x=\lambda_y=0.0007$, (b) $\lambda_{x}=0.0022$ and
$\lambda_{y}=0.0052$, and (c) $\lambda_{x}=0.0052$ and $\lambda_{y}=-0.0022$.
All plots depict the $5\times 5$ sublattice around the breather.  
Dissipative breathers are very narrow and essentially confined on one lattice site.
}
\label{Fig4}
\end{figure}

Localized modes in the damped-driven case are constructed for $\gamma=0.01$, $\varepsilon_0=0.04$
and $\alpha=+1$ with the  method  described in Ref.~\cite{LET}.
The resulting localized modes are called {\em dissipative breathers}, and their examples are shown in
Fig.~\ref{Fig4} for $T_{b}=5.8$ and (a) $\lambda_x=\lambda_y=0.0007$,
(b) $\lambda_{x}=0.0022$ and $\lambda_{y}=0.0052$,  and
(c) $\lambda_{x}=0.0052$ and $\lambda_{y}=-0.0022$. The dissipative
modes have been evolved in time, and we found that at long times some dissipative
breathers constructed for relatively large couplings loose their initial shape and finally decay.

\begin{figure}[!ht]
\includegraphics[angle=0, width=1. \linewidth]{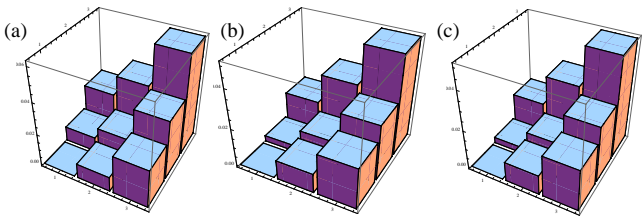}
\caption{Histogram of the breather Hamiltonian. Breather difference energies 
$\Delta E$ for $\alpha=+1$, $Tb=6.69$ constructed in the upper left 
$3\times 3$ corner of the lattice. Case (a) $\lambda_x=\lambda_y=0.028$, case 
(b) $\lambda_x=0.026$ and $\lambda_y=0.029$, and case (c) $\lambda_x=0.029$ 
and $\lambda_y=-0.026$. To evaluate  $\Delta E$ we calculate the energy of the 
breathers centered at different sites and subtract the energy of the corner 
breather.
}
\label{Fig5}
\end{figure}

Additionally, we calculate the total energy of discrete breathers in a 
lattice with planar and planar-axial configuration for $\alpha=+1$ and 
$T_{b}=6.69$ (Hamiltonian case). Figure~\ref{Fig5} shows the energy histograms 
of the relevant corner of the lattice normalized to the energy of the corner 
$(1,1)$ breather. In order to construct the histograms centered in each of 
the lattice sites, we  normalized it to the edge breather energy. In the case 
(a) the discrete breather is constructed in a lattice of coupling 
$\lambda_x=\lambda_y=0.028$, in (b) the case with anisotropy in couplings 
$\lambda_x=0.026$ and $\lambda_y=0.029$,
while in the case (c), couplings are $\lambda_x=0.029$ and $\lambda_y=-0.026$.
The energy of the discrete breathers as a function of the lattice site 
increases, i.e, as the discrete breather is constructed in the interior of 
the lattice energy is larger compared to the discrete breather that is 
located in the corner of the lattice. 

\begin{figure}[!ht]
\includegraphics[angle=0, width=1. \linewidth]{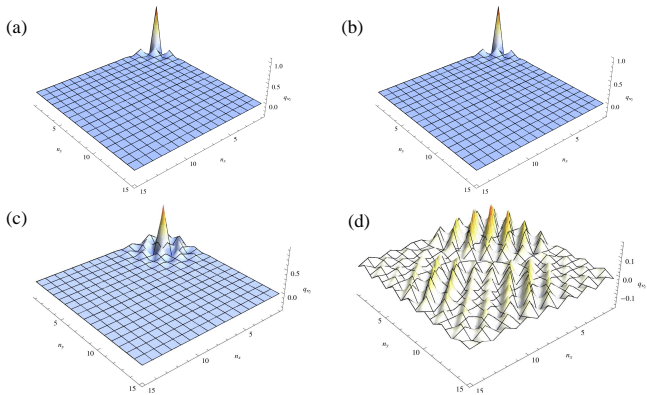}
\caption{Amplitudes $q_{n,m}$ of the breather for $\alpha=+1$, $T_{b}=6.69$ and
$\lambda_x=\lambda_y=0.028$, constructed on the site (1,1) for (a) t=0  and 
(b) $t=1450T_{b}$, and the breather constructed on site (3,3), for (c) $t=0$ 
and (d) $t=1450T_{b}$.
}
\label{Fig6}
\end{figure}

We note that in the one-dimensional 
case the bulk breathers have lower energy compared to the surface 
ones~\cite{LTK} while in two-dimensional lattice the behavior is the 
contrary.  We thus find that two-dimensional surface and especially edge 
breathers form easier.

Finally, we study the time evolution of the discrete breather that is 
constructed in the corner site (1,1)
and compare this case with a discrete breather centered at the (3,3) site
for the coupling $\lambda_x=\lambda_y=0.028$. The breather of the
latter case after $t=95 T_{b}$ starts to loose its shape, in contrast to
the breather of (1,1) site which survives for much longer times, viz.  
$t=1450T_{b}$ [see Fig.~(\ref{Fig6})].  For different coupling values such as 
$\lambda_x=\lambda_y=0.01$ we find that both the corner (1,1) and inner (3,3) 
breathers remain stable for at least $t=1450T_{b}$.  This feature, while 
compatible with the fact that the corner breathers are more stable than inner 
ones, shows additionally that in finite lattices small changes in parameters
may affect the stability properties of the breathers~\cite{Morgante}.

In conclusion, we have studied surface discrete breathers located in the 
corner and at the edge of the two-dimensional lattices of the split-ring 
resonators. Using standard numerical methods, we have found nonlinear 
localized modes both in the Hamiltonian and dissipative systems. 
Two-dimensional breathers in conservative lattices have been found to be 
linearly stable for up to certain (large) values of the coupling coefficient, 
in both planar and planar-axial configurations of the split-ring-resonator 
lattices. Dissipative discrete surface breather can retain their shapes for 
several periods of time, and they depending critically on the lattice coupling. Finally, we have found that the discrete breathers located deep inside 
the lattice have higher energy compared to the breathers located in the 
corners and at the edges. This distinct two-dimensional feature of nonlinear 
localized modes contrasts with the one-dimensional behavior being attributed 
to the larger number of neighbors of the two-dimensional lattice. Furthermore, 
the two-dimensional breathers located inside the lattice loose rapidly their
initial shape as they evolve in time while the surface breathers are seen to 
be stable at least for  $t \approx 1500T_{b}$.

\end{document}